\documentclass[twocolumn,amsmath,amssymb,prl,superscriptaddress]{revtex4-2}

\usepackage{graphicx}
\usepackage{color}
\usepackage{tabularx}
\usepackage{multirow}
\usepackage{booktabs}
\usepackage{gensymb}
\usepackage[T1]{fontenc}
\usepackage{hyperref}

\begin{document}

\newcommand*{\TUD}{Technische Universit\"at Darmstadt, Fachbereich Physik, 64289 Darmstadt, Germany}
\newcommand*{\GSI}{GSI Helmholtzzentrum f\"ur Schwerionenforschung GmbH, Planckstr. 1, 64291 Darmstadt, Germany}
\newcommand*{\FAIR}{Helmholtz Forschungsakademie Hessen f\"ur FAIR, Max-von-Laue-Str. 12, 60438 Frankfurt, Germany}
\newcommand*{\CAEN}{Université de Caen Normandie, ENSICAEN, IN2P3/CNRS, LPC-Caen UMR6534, F-14000 Caen, France}
\newcommand*{\RIKEN}{RIKEN Nishina Center for Accelerator-Based Science, 2-1 Hirosawa, Wako 351-0198, Japan}
\newcommand*{\TEXAS}{Department of Physics and Astronomy, East Texas A$\&$M University, Commerce, TX 75429, USA}
\newcommand*{\TUM}{Technische Universit\"at M\"unchen, Physik Department, 85748 Garching, Germany}
\newcommand*{\CEA}{IRFU, CEA, Universit\'{e} Paris-Saclay, F-91191 Gif-sur-Yvette, France}
\newcommand*{\SANT}{Dpt. de F\'{i}sica de Part\'{i}culas, Universidade de Santiago de Compostela, E-15782 Santiago de Compostela, Spain}
\newcommand*{\KVI}{Nuclear Energy group, ESRIG, University of Groningen, 9747 AA Groningen, The Netherlands}
\newcommand*{\ELKH}{HUN-REN ATOMKI, P.O. Box 51, H-4001 Debrecen, Hungary}
\newcommand*{\HUN}{Institute of Physics, Faculty of Science and Technology, University of Debrecen, Egyetem tér 1, H-4032 Debrecen, Hungary}
\newcommand*{\PEK}{State Key Laboratory of Nuclear Physics and Technology, School of Physics, Peking University, Beijing 100871, China}
\newcommand*{\YORK}{School of Physics, Engineering and Technology, University of York, York YO10 5DD}
\newcommand*{\RBI}{Rudjer Bo\v{s}kovi\'{c} Institute, Zagreb, Croatia}
\newcommand*{\TITECH}{Department of Physics, Institute of Science Tokyo, 2-12-1 O-Okayama, Meguro, Tokyo 152-8551, Japan}
\newcommand*{\Seoul}{Department of Physics and Astronomy, Seoul National University, 599 Gwanak, Seoul 151-742, Republic of Korea}
\newcommand*{\Tohoku}{Department of Physics, Tohoku University, Miyagi 980-8578, Japan}
\newcommand*{\HKU}{Department of Physics, The University of Hong Kong, Hong Kong, China}
\newcommand*{\FUD}{	Key Laboratory of Nuclear Physics and Ion-Beam Application (MOE), Institute of Modern Physics, Fudan University, Shanghai 200433, China}
\newcommand*{\GOTHE}{Department of Physics, Chalmers University of Technology, 412 96, G{\"o}teborg, Sweden}
\newcommand*{\CNS}{Center for Nuclear Study, The University of Tokyo, 7-3-1 Hongo, Bunkyo, Tokyo 113-0033, Japan}
\newcommand*{\KOLN}{Universit{\"a}t zu K{\"o}ln, Institut f{\"u}r Kernphysik, Z{\"u}lpicher Stra\ss e 77, 50937 K{\"o}ln, Germany}
\newcommand*{\GANIL}{GANIL, CEA/DRF-CNRS/IN2P3, 14076 Caen, France}
\newcommand*{\ISC}{Institute of Space Sciences, 077125 Magurele, Romania}
\newcommand*{\TOK}{University of Tokyo, Japan}
\newcommand*{\EWHA}{Ewha Womans University, Seoul 03760, Korea}
\newcommand*{\IBS}{Center for Exotic Nuclear Studies, Institute for Basic Science, Daejeon 34126, Korea}
\newcommand*{\Kyoto}{Department of Physics, Kyoto University, Kitashirakawa-Oiwake, Sakyo, Kyoto 606-8502, Japan}
\newcommand*{\MAINZ}{Institut f\"ur Kernphysik and PRISMA+ Cluster of Excellence, Johannes Gutenberg-Universit\"at, 55128 Mainz, Germany}
\newcommand*{\FRIB}{Facility for Rare Isotope Beams, Michigan State University, East Lansing, MI 48824, USA}
\newcommand*{\ORNL}{Physics Division, Oak Ridge National Laboratory, Oak Ridge, TN 37831, USA}
\newcommand*{\MSU}{Department of Physics and Astronomy, Michigan State University, East Lansing, MI 48824, USA}

\def\Label{}

\title{The dipole strength distribution of $^8$He and decay characteristics}

\author{C.~Lehr}
\affiliation{\TUD}

\author{T.~Aumann}
\affiliation{\TUD}
\affiliation{\GSI}
\affiliation{\FAIR}

\author{M.~Duer}
\email[Corresponding Author:\ ]{Meytal Duer (mduer@ikp.tu-darmstadt.de)}
\affiliation{\TUD}

\author{A.~T.~Saito}
\affiliation{\TITECH}

\author{T.~Nakamura}
\affiliation{\TITECH}

\author{N.~L.~Achouri}
\affiliation{\CAEN}

\author{D.~Ahn}
\affiliation{\RIKEN}

\author{H.~Baba}
\affiliation{\RIKEN}

\author{S.~Bacca}
\affiliation{\MAINZ}

\author{C.~A.~Bertulani}
\affiliation{\TEXAS}

\author{M.~B\"ohmer}
\affiliation{\TUM}

\author{F.~Bonaiti}
\affiliation{\MAINZ}
\affiliation{\FRIB}
\affiliation{\ORNL}

\author{K.~Boretzky}
\affiliation{\GSI}

\author{C.~Caesar}
\affiliation{\TUD}
\affiliation{\GSI}
\affiliation{\RIKEN}

\author{N.~Chiga}
\affiliation{\RIKEN}

\author{D.~Cortina-Gil}
\affiliation{\SANT}

\author{C.~A.~Douma}
\affiliation{\KVI}

\author{F.~Dufter}
\affiliation{\TUM}

\author{Z.~Elekes}
\affiliation{\ELKH}
\affiliation{\HUN}

\author{J.~Feng}
\affiliation{\PEK}

\author{B.~Fern{\'a}ndez-Dom\'{\i}nguez}
\affiliation{\SANT}

\author{U.~Forsberg}
\affiliation{\YORK}

\author{N.~Fukuda}
\affiliation{\RIKEN}

\author{I.~Gasparic}
\affiliation{\RBI}
\affiliation{\TUD}
\affiliation{\RIKEN}

\author{Z.~Ge}
\affiliation{\RIKEN}

\author{R.~Gernh{\"a}user}
\affiliation{\TUM}

\author{J.~M.~Gheller}
\affiliation{\CEA}

\author{J.~Gibelin}
\affiliation{\CAEN}

\author{A.~Gillibert}
\affiliation{\CEA}

\author{K.~I.~Hahn}
\affiliation{\EWHA}
\affiliation{\IBS}

\author{Z.~Hal\'{a}sz}
\affiliation{\ELKH}

\author{M.~N.~Harakeh}
\affiliation{\KVI}

\author{A.~Hirayama}
\affiliation{\TITECH}

\author{M.~Holl}
\affiliation{\TUD}

\author{N.~Inabe}
\affiliation{\RIKEN}

\author{T.~Isobe}
\affiliation{\RIKEN}

\author{J.~Kahlbow}
\affiliation{\TUD}

\author{N.~Kalantar-Nayestanaki}
\affiliation{\KVI}

\author{D.~Kim}
\affiliation{\IBS}

\author{S.~Kim}
\affiliation{\IBS}
\affiliation{\TUD}

\author{T.~Kobayashi}
\thanks{Deceased}
\affiliation{\Tohoku}

\author{Y.~Kondo}
\affiliation{\TITECH}

\author{D.~K{\"o}rper}
\affiliation{\GSI}

\author{P.~Koseoglou}
\affiliation{\TUD}

\author{Y.~Kubota}
\affiliation{\RIKEN}

\author{P.~J.~Li}
\affiliation{\HKU}
\affiliation{\FUD}

\author{S.~Lindberg}
\affiliation{\GOTHE}

\author{Y.~Liu}
\affiliation{\PEK}

\author{F.~M.~Marqu\'es}
\affiliation{\CAEN}

\author{S.~Masuoka}
\affiliation{\CNS}

\author{M.~Matsumoto}
\affiliation{\TITECH}

\author{J.~Mayer}
\affiliation{\KOLN}

\author{K.~Miki}
\affiliation{\TUD}
\affiliation{\Tohoku}

\author{M.~Miwa}
\affiliation{\RIKEN}

\author{B.~Monteagudo}
\affiliation{\CAEN}

\author{A.~Obertelli}
\affiliation{\TUD}
\affiliation{\CEA}

\author{N.~A.~Orr}
\affiliation{\CAEN}

\author{H.~Otsu}
\affiliation{\RIKEN}

\author{V.~Panin}
\affiliation{\RIKEN}
\affiliation{\GSI}

\author{S.~Y.~Park}
\affiliation{\EWHA}
\affiliation{\IBS}

\author{M.~Parlog}
\affiliation{\CAEN}

\author{S.~Paschalis}
\affiliation{\YORK}
\affiliation{\TUD}

\author{P.~M.~Potlog}
\affiliation{\ISC}

\author{S.~Reichert}
\affiliation{\TUM}

\author{A.~Revel}
\affiliation{\CAEN}
\affiliation{\GANIL}
\affiliation{\FRIB}
\affiliation{\MSU}

\author{D.~M.~Rossi}
\affiliation{\TUD}

\author{R.~Roth}
\affiliation{\TUD}

\author{M.~Sasano}
\affiliation{\RIKEN}

\author{H.~Scheit}
\affiliation{\TUD}

\author{F.~Schindler}
\affiliation{\TUD}

\author{T.~Shimada}
\affiliation{\TITECH}

\author{S.~Shimoura}
\affiliation{\CNS}

\author{H.~Simon}
\affiliation{\GSI}

\author{S.~Storck~Dutine}
\affiliation{\TUD}

\author{L.~Stuhl}
\affiliation{\IBS}
\affiliation{\CNS}

\author{H.~Suzuki}
\affiliation{\RIKEN}

\author{D.~Symochko}
\affiliation{\TUD}

\author{H.~Takeda}
\affiliation{\RIKEN}

\author{S.~Takeuchi}
\affiliation{\TITECH}

\author{J. Tanaka}
\affiliation{\TUD}
\affiliation{\RIKEN}

\author{Y.~Togano}
\affiliation{\TITECH}

\author{T.~Tomai}
\affiliation{\TITECH}

\author{H.~T.~T{\"o}rnqvist}
\affiliation{\TUD}
\affiliation{\GSI}

\author{J.~Tscheuschner}
\affiliation{\TUD}

\author{T.~Uesaka}
\affiliation{\RIKEN}

\author{V.~Wagner}
\affiliation{\TUD}

\author{H.~Yamada}
\affiliation{\TITECH}

\author{B.~Yang}
\affiliation{\PEK}

\author{L.~Yang}
\affiliation{\CNS}

\author{Z.~H.~Yang}
\affiliation{\RIKEN}

\author{M.~Yasuda}
\affiliation{\TITECH}

\author{K.~Yoneda}
\affiliation{\RIKEN}

\author{L.~Zanetti}
\affiliation{\TUD}

\author{J.~Zenihiro}
\affiliation{\RIKEN}
\affiliation{\Kyoto}

\date{\today}

\begin{abstract}
The weak binding and spatially extended neutron densities characteristic of drip-line nuclei give rise to a distinctive low-energy dipole response. The drip-line nucleus $^8$He is the most neutron-rich bound nucleus with a mass-to-charge ratio of $A/Z=4$. 
We measure the dipole response of $^8$He, including for the first time the four-neutron decay channel. A total dipole strength of $\sum B(E1)(E^*<15$~MeV$)=0.95(16)~e^2$fm$^2$ and a dipole polarizability of $\alpha_D = 0.61(1)$~fm$^3$ are extracted from the differential Coulomb-excitation cross section and compared to state-of-the-art theoretical calculations employing coupled cluster and three-body approaches. We find that the dipole continuum is dominated, even at high excitation energies well above the $4n$ decay threshold, by two-neutron emission, pointing to a $^6$He$+2n$ structure of the excited dipole mode. No indication was found for a $4n$ final-state correlation, while pronounced $nn$ and $^6$He-$n$ final-state correlations are apparent.
\end{abstract}

\maketitle

{\it Introduction}.--- 
The dipole strength distributions of nuclei are a cornerstone of nuclear physics~\cite{Aumann2013,BaccaPastore,Harakeh2001}. They are essential for applications in nuclear astrophysics~\cite{Arnould_2007}, medical isotope production~\cite{wang22}, and fission~\cite{chadwick11}. At the same time, they provide key insights into nuclear structure and dynamics, especially at the limits of stability.  
At the neutron drip line, the weak binding of excess neutrons can lead to intriguing phenomena such as the development of neutron halos and thick neutron skins. These structural changes manifest in a characteristic manner in the electric dipole ($E1$) response. For stable nuclei, the $E1$ strength is dominated by the isovector giant dipole resonance (IVGDR), commonly understood as a collective harmonic vibration of neutrons versus protons. A unique feature of exotic nuclei is the appearance of significant $E1$ strength at lower excitation energies, known as soft $E1$ excitation, which is a direct consequence of the extended neutron densities. The mechanism of an enhanced $E1$ strength can be of non-resonant character, $i.e.$, a direct breakup leading to continuum states~\cite{Hansen1987}, as established for 1$n$ halos, $e.g.$, $^{11}$Be~\cite{Nakamura1994}, or a resonance occurring as an oscillation of the core against the halo neutron(s)~\cite{Ikeda1992}. In 2$n$ halos, $e.g.$, $^6$He~\cite{Aumann1999,Sun2021} and $^{11}$Li~\cite{Nakamura2006}, the response is currently also understood as attributed to a non-resonant origin, while it is sensitive to ground-state correlations among the neutrons, as well as final-state interactions. Of particular interest is $^8$He, the most neutron-rich bound nucleus, with an extreme mass-to-charge ratio of $A/Z=4$, where the four excess neutrons are considered as forming a neutron-skin structure~\cite{Tanihata1992}.

From the theoretical side, while the description of collective excitations has been traditionally dominated by density functional theory (DFT), especially in heavy-mass nuclei, first ab initio calculations recently became available for light- and medium-mass nuclei.
A study of the $^8$He dipole properties was carried out using coupled-cluster (CC) theory combined with the Lorentz integral transform (LIT)~\cite{Bonaiti2022} in a LIT-CC approach, revealing a low-energy dipole strength around 
$E^*$ = 5~MeV. Another study employing the importance-truncated no-core shell model (IT-NCSM)~\cite{Stumpf2018} likewise found a significant strength at low excitation energies. This is in contrast, $e.g.$, with  a recent  calculation  based on the random-phase approximation (RPA) within the DFT framework~\cite{Piekarewicz2022}, which disfavored the emergence of a soft dipole mode in $^8$He, attributing the low-energy feature in the calculation to a spurious center of mass contamination. Experimental data can shed light on this controversy.

An established tool to experimentally study the $E1$ response of exotic nuclei is relativistic Coulomb excitation. In such a reaction, the projectile is excited by the Lorentz-contracted electromagnetic field of a high-$Z$ target. If it is excited beyond its particle decay threshold, it will subsequently breakup. In $^8$He, the lowest particle emission threshold is $^6$He+2$n$ with a two-neutron separation energy of $S_{2n}$ = 2.1~MeV, while the four-neutron channel opens at an excitation energy of $S_{4n}$ = 3.1~MeV. Coulomb excitation data for $^8$He are rather scarce, and in particular, only the two-neutron decay channel has been measured so far. A measurement performed at GSI at an incident beam energy of 227~MeV/nucleon~\cite{Meister2002}, indicated the existence of a soft dipole resonance at $E^*\sim$~4~MeV. It was further supported via a transfer reaction measurement~\cite{Gologov2009}, although at a lower excitation energy of 3~MeV. On the other hand, an experiment carried out at MSU with a beam energy of 24~MeV/nucleon~\cite{Iwata2000}, attributed an insignificant fraction (<3\%) of the total energy-weighted dipole sum rule to a possible soft dipole mode. A recent inelastic proton-scattering experiment~\cite{Holl2021} confirmed this result, where the measured angular distribution was shown to be inconsistent with a dipole excitation. As both Coulomb dissociation experiments feature low statistics and measured only the $^6$He+2$n$ decay channel, a meaningful interpretation of the data and comparison to ab initio theory was difficult.  

In this work, the high beam intensity provided at the Radioactive Ion Beam Factory (RIBF) in combination with a dedicated neutron detection setup, allowed to obtain a high-statistics data set with an extended excitation-energy range (up to 15~MeV) and to measure for the first time the $^4$He+4$n$ decay channel. Detection of four neutrons in coincidence is a challenging task, and only very recently the first successful measurement was demonstrated~\cite{Kondo2023}, within the same experimental campaign.

{\it Experiment}.--- The experiment was performed at the RIBF operated by the RIKEN Nishina Center and the Center for Nuclear Study, University of Tokyo. The secondary beam of $^8$He was produced at an energy of $\sim$185~MeV/nucleon by fragmentation of a primary $^{18}$O beam on a beryllium target, and selected using the BigRIPS fragment separator~\cite{Kubo2007}. Projectiles were identified on an event-by-event basis via measurement of their energy deposition and time-of-flight (TOF) in thin plastic scintillators and transported to the SAMURAI spectrometer setup~\cite{Kobayashi2013} with a typical $^8$He intensity of 1.1$\cdot$10$^5$ particles per second with 98\% purity. Using two drift chambers, the projectiles were tracked towards the solid targets, where nuclear and Coulomb-excitation processes were induced. The main data were taken with a 3.157(47)~g/cm$^2$ lead target. In addition, a set of targets with increasing $Z$ was used, namely, carbon (2.371(25)~g/cm$^2$), titanium (3.115(49)~g/cm$^2$), and tin (2.922(47)~g/cm$^2$). Data from these targets were used to correct for background from nuclear excitation reactions with high accuracy. An empty-target setting was used for additional background subtraction.  
The multi-neutron decay of $^8$He following electromagnetic excitation was measured in complete kinematics via the $^6$He+2$n$ and $^4$He+4$n$ channels. The SAMURAI spectrometer, operated with a central magnetic field of 2.6~T, provided momentum analysis of the outgoing charged particles. Their trajectories were determined using two drift chambers, placed before and after the dipole magnet, while scintillator walls at the focal plane were used for energy deposition and TOF measurements. Decay neutrons were measured at forward angles by three large-area segmented scintillator walls, consisting of the NeuLAND demonstrator~\cite{Boretzky2021} and the NEBULA array~\cite{Nakamura2016,Kondo2020}. The NeuLAND demonstrator, with an overall depth of 40~cm, was placed $\sim$10.5~m downstream the target and separated by $\sim$2~m from NEBULA. The latter was arranged in a two-wall configuration separated by 85~cm, with a depth of 2$\times$24~cm. An additional thin plane of scintillators was placed in front of each wall to veto background from charged particles. The neutron momenta were derived from the TOF and the hit position measured in the detector arrays. A dedicated multi-neutron reconstruction algorithm based on Geant4~\cite{Agostinelli2003} simulations was developed, with special care to cross-talk rejection, in particular for the 4$n$ case~\cite{lehr2021}. This resulted in total detection efficiencies of $\sim$1.2\% for the $^4$He+4$n$ channel, and $\sim$12\% for $^6$He+2$n$ at 1~MeV relative energy of the 5-body and 3-body systems, respectively. The simulations were validated as part of the experimental campaign by comparing the single-neutron detection efficiency~\cite{Kondo2020}. To account for small deviations, a scaling factor is applied with an associated systematic uncertainty ($\sim4\%$). For the cross-talk contribution, a 10\% relative uncertainty is estimated.
The decay energy of $^8$He is reconstructed from the relative energy of the decay products ($E_{fxn}$) based on the invariant-mass method, using the four-momenta of the fragment ($f$) and $x$ neutrons. Simulated resolutions of 0.2 and 0.24~MeV ($\sigma$) were obtained at 1~MeV relative energy for the two channels, respectively.    

\begin{figure}[htbp!]
	\includegraphics[width=\linewidth]{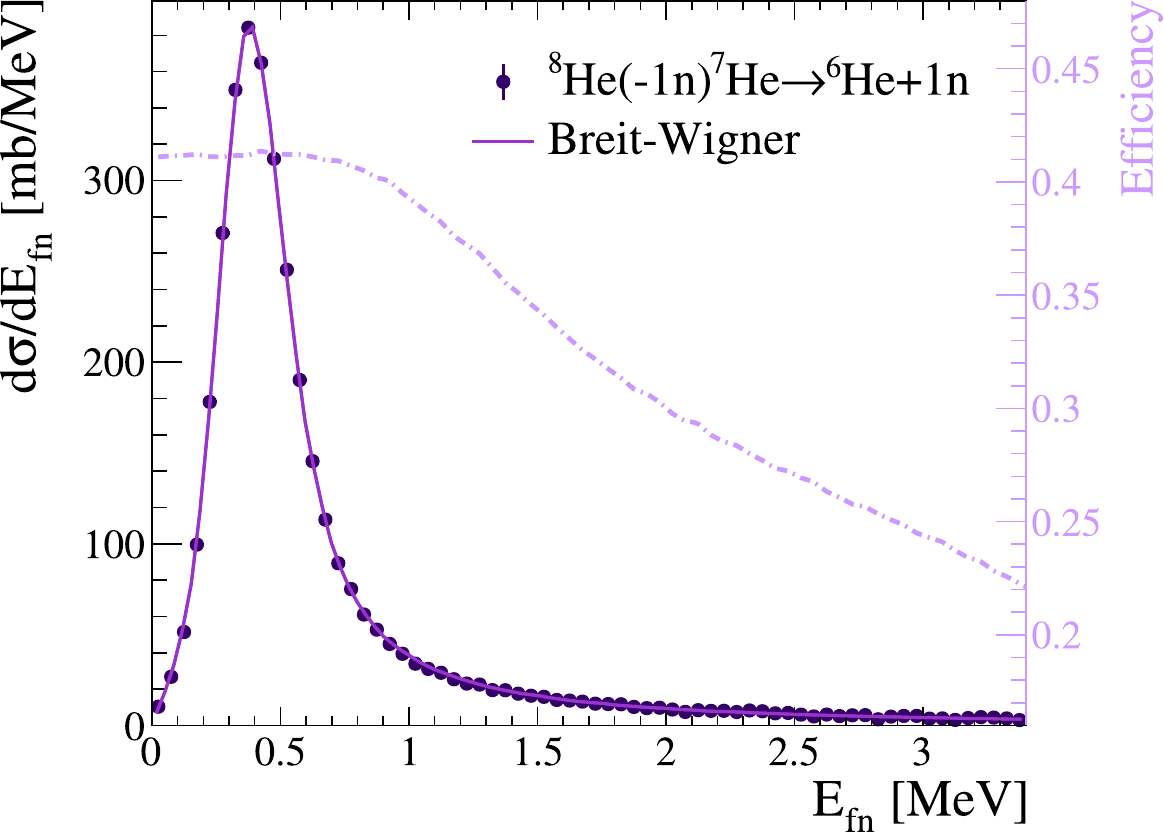}
	\caption{Two-body relative-energy spectrum of the $^7$He resonance populated by $^8$He(-1$n$) removal reaction induced by a carbon target. The dark purple curve represents a fit to an energy-dependent Breit-Wigner line shape, convoluted with the experimental response. The dotted-dashed curve displays the overall detection efficiency.	}	
\label{fig:7He}
\end{figure}

To demonstrate the neutron detection performance, the two-body decay of the $^7$He ground-state resonance into $^6$He+1$n$ has been measured as a benchmark, populated via the one-neutron removal reaction $^8$He(-1$n$) with a carbon target. 
The corresponding relative-energy ($E_{fn}$) spectrum was extracted and corrected for the detection efficiency to deduce the cross section. Following that, background contributions from empty-target data as well as remaining spurious events from the 2$n$ breakup channels, which can be misidentified as 1$n$ events, were subtracted. The $^7$He spectrum is shown in Fig.~\ref{fig:7He}, exhibiting the known 3/2$^-$ ground-state resonance at $E_{fn}\sim0.4$~MeV. The spectrum is fitted with an energy-dependent Breit-Wigner line shape convoluted with the experimental response to account for the relative-energy resolution ($\sim$90~keV ($\sigma$) at $E_{fn}$=0.4~MeV). The resonance energy and width, determined from the fit are $E_{\rm r} = 385\pm1$(stat.)$\pm$1(sys.)~keV and $\Gamma = 193\pm1$(stat.)$\pm$4(sys.)~keV, respectively. These are in very good agreement with previously measured values using liquid-hydrogen target~\cite{Aksyutina2009,Denby2008,Renzi2016}, but have been determined here with higher precision. The integrated cross section ($E_{fn}<5$~MeV) for populating the $^7$He resonant state is extracted as $185\pm1$(stat.)$\pm$15(sys.)~mb.

{\it Results}.--- The $^6$He+2$n$ and $^4$He+4$n$ breakup channels of $^8$He induced by a lead target were identified by selecting the outgoing fragment in coincidence with 2 and 4 neutrons, respectively. The three- and five-body relative-energy spectra are then reconstructed via the invariant mass following subtraction of background contributions as determined from the empty-target measurement. For cross-talk correction, the corresponding contributions for the two channels are considered and subtracted (see Supplemental Material~\cite{Supp}). Finally, the measured spectra are corrected for detection efficiency to determine the differential cross section.  

To extract pure electromagnetic properties and the dipole response of $^8$He, the nuclear part of the cross section has to be eliminated. This is often done by a measurement with a light target such as carbon~\cite{Aumann2019}, where the electromagnetic excitation cross section is small. Under the assumption that the nuclear spectra have the same shape when impinging on both target nuclei, the obtained cross section is then scaled and subtracted from the total cross section measured with the lead target. The scaling factor is typically obtained semi-empirically~\cite{Rossi2013}, or by a theoretical calculation~\cite{Aumann1999}. In this work, it was determined independently based on measurements with a series of targets, thus minimizing model dependence. While the Coulomb excitation cross section scales approximately as $Z^2$, the nuclear cross section for peripheral reactions relevant for halo nuclei scales as $A^{1/3}$~\cite{Aumann2019}. This dependency is used to extract the scaling factor from the data by assuring that the extracted $B(E1)$ distribution is independent from the target (Sn, Pb) by introducing a free parameter for the scaling and applying a fitting procedure. Finally, the small electromagnetic contribution within the carbon-target spectrum is corrected for (see Supplemental Material~\cite{Supp}).

\begin{figure}[htbp!]
	\includegraphics[width=\linewidth]{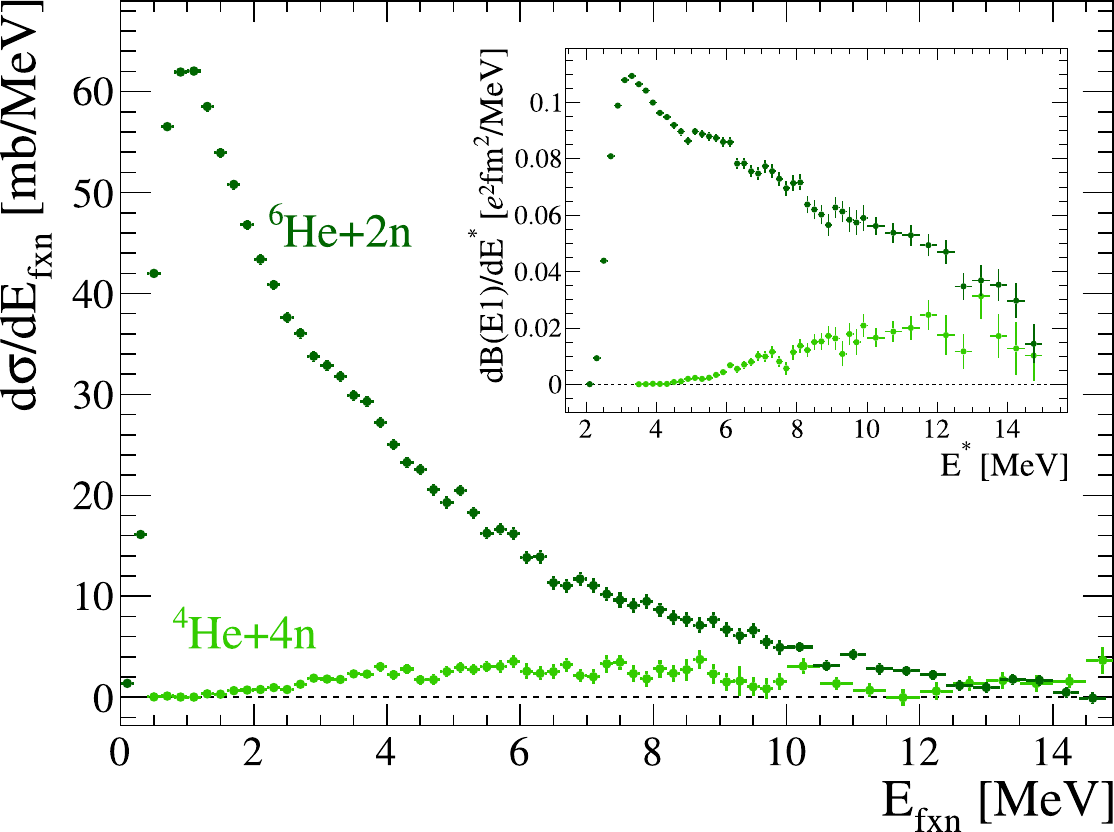}
	\caption{Coulomb excitation cross section spectra for three- (dark green) and five-body (light green) breakup channels of $^8$He measured with lead target. The inset shows the corresponding dipole strength distributions, $dB(E1)/dE^*$, as a function of the excitation energy.	}	
\label{fig:2n4n}
\end{figure}

The differential Coulomb excitation cross sections for the 2$n$ and 4$n$ channels are shown in Fig.~\ref{fig:2n4n}. The integrated cross sections up to $E^*$ = 15~MeV are $\sigma_{C,2n}=243\pm1({\rm stat.})\pm33({\rm sys.})$~mb and $\sigma_{C,4n}=21.14\pm0.02({\rm stat.})\pm5.34({\rm sys.})$~mb. Systematic errors include uncertainties of the detection efficiency, target areal densities and the scaling factor for the subtraction of the nuclear contribution. From the differential cross section, the dipole strength distributions $dB(E1)/dE^*$ are derived following Ref.~\cite{Bertulani1988}. The excitation process is treated as a photo-absorption induced by a virtual photon, known as the equivalent-photon method. The Coulomb breakup cross section for electric dipole excitation is expressed as
\begin{equation}
    \frac{d\sigma(E1)}{dE^*} = \frac{16\pi^3}{9\hbar c}N_{E1}(E^*)\frac{dB(E1)}{dE^*},
\end{equation}
where $N_{E1}(E^*)$ is the number of virtual photons at excitation energy $E^*$. This relation connects the measured cross section to the reduced transition probability $B(E1)$, which contains the physics information of the dipole response (see Fig.~\ref{fig:2n4n} inset). It can be clearly observed that the 2$n$ channel is dominant even at high excitation energies well above the 4$n$ threshold. This surprising result points towards a characteristic structure effect of the $^8$He dipole continuum. The dipole vibration seems to be dominated by a $^6$He-di-neutron component, which would explain the preferred decay into $^6$He+2$n$.

\begin{figure}[t!]
	\includegraphics[width=\linewidth]{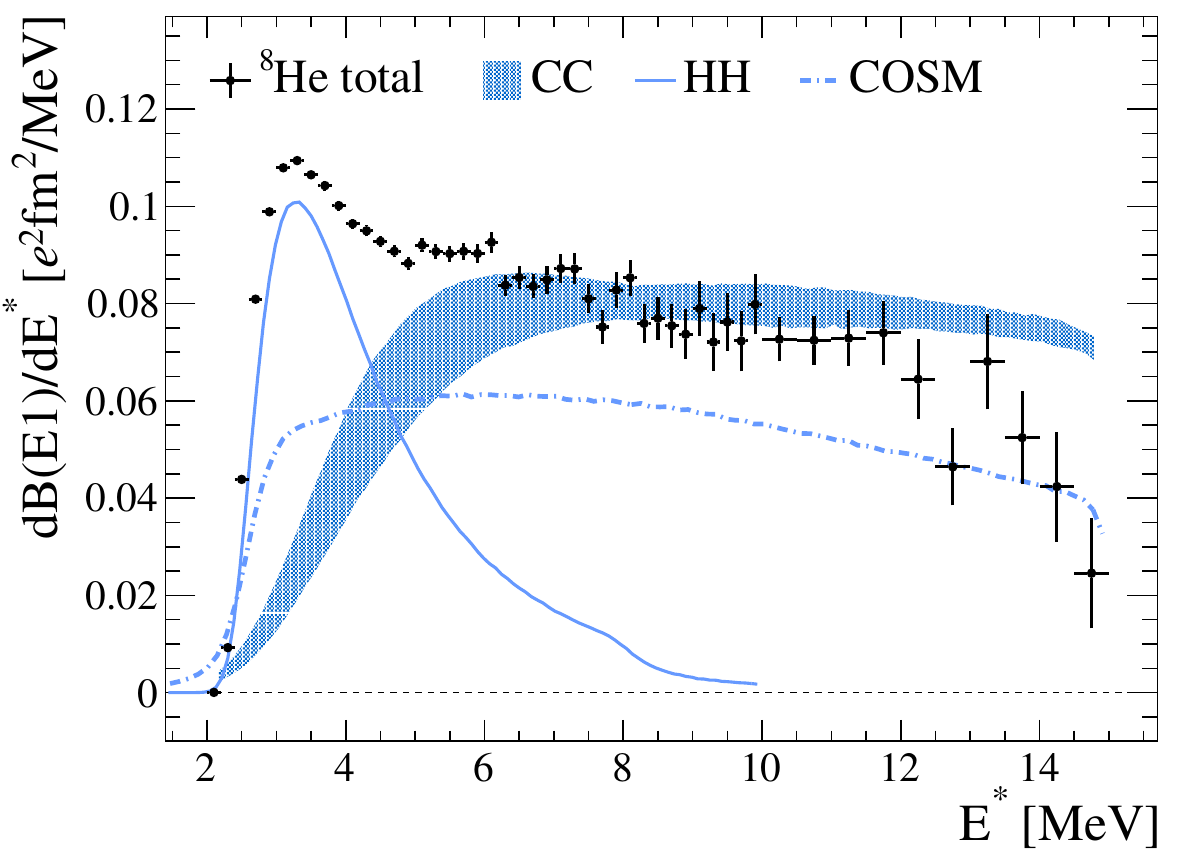}
	\caption{ Total $B(E1)$ spectrum measured with lead target. Theory predictions folded with the experimental response are shown in blue based on CC~\cite{Bonaiti2022} (band), HH~\cite{Grigorenko2009} (solid curve) and COSM~\cite{Myo2022} (dotted-dashed curve).   } 
\label{fig:BE1}
\end{figure}

The total $dB(E1)/dE^*$ spectrum, (sum of the 2$n$ and 4$n$ channels) is shown in Fig.~\ref{fig:BE1}. From that, the non-energy-weighted dipole strength integrated up to $E^*$=15~MeV amounts to $\Sigma B(E1)=0.95\pm0.01({\rm stat.})\pm0.16({\rm sys.})~e^2{\rm fm}^2$ and the energy-weighted one to $\Sigma E^* B(E1)=7.55\pm0.18({\rm stat.})\pm1.39({\rm sys.})~e^2{\rm fm}^2$MeV, which corresponds to about 1/3 of the Thomas-Reiche-Kuhn sum rule~\cite{Sagawa1990}. With the inverse energy-weighted sum of the dipole strength, the electric dipole polarizability is extracted as
\begin{equation}
    \alpha_D = \frac{8\pi}{9} \int \frac{dB(E1)/dE^*}{E^*}dE^*,
\end{equation}
resulting in $\alpha_D=0.61\pm0.01({\rm stat.})\pm0.01({\rm sys.})~{\rm fm}^3$. 

The $B(E1)$ distribution exhibits a clear soft-dipole enhancement at 3~MeV. This is consistent with previous experiments~\cite{Meister2002,Gologov2009} where only the 2$n$ channel was measured. The weak contribution of the $4n$ channel to the total strength confirms it is not caused by this channel opening at around this energy, thus resolving discussions on the interpretation of the peak observed in earlier experiments.

We compare the data with ab initio predictions from the LIT-CC method~\cite{Bonaiti2022,Bonaiti2024}. Refs.~\cite{Bonaiti2022,Bonaiti2024} focused on moments of the dipole response function of $^{8}$He, and in particular on $\alpha_D$, which was found to vary between $0.37(3)$ to $0.48(3)$ fm$^3$ employing different chiral effective field theory interactions. In the present work, we determine the full dipole response function, $dB(E1)/dE^*$, by numerically inverting the Lorentz integral transform. We focus on the chiral  interaction 1.8/2.0 (EM)~\cite{Hebeler2011}, which provides the largest $\alpha_D$. LIT-CC calculations are performed with the best approximation scheme at disposal, labeled CCSDT-1, including linear triples corrections in both the ground- and excited-state calculations~\cite{miorelli2018}. The resulting response function is shown in Fig.~\ref{fig:BE1} with a blue band quantifying the uncertainties associated with the inversion procedure and model-space convergence. Further details on the LIT-CC calculations are provided in the Supplemental Material~\cite{Supp}. While it describes the strength distribution rather well at excitation energies above 5-6~MeV, it does not reproduce the low-energy part of the spectrum which peaks at $\sim 3$~MeV. 
A possible origin for this discrepancy may lie in missing higher-order correlations within the CC many-body expansion. To explore this, we compare LIT-CC calculations with results from the IT-NCSM approach~\cite{Stumpf2018}, which is expected to capture a larger portion of many-body correlations for the same interaction. Similar to the LIT-CC case, the spread in $\alpha_D$ predictions for different chiral forces remains significant, ranging from 0.4454(19) to 0.5950(1)~fm$^3$. For the NNLO$_{\rm sat}$ interaction~\cite{ekstrom2015}, where both results are available, IT-NCSM predicts $\alpha_D = 0.4454(19)$~fm$^3$, which is more than 10\% higher than the corresponding LIT-CC value of $0.37(3)$~fm$^3$, suggesting that many-body correlations beyond triples might play a role for this nucleus. Since only a discrete strength distribution is calculated in this case, it cannot be compared directly to the data.

The strength function for the 2$n$ breakup channel was also estimated in a three-body model and solving the Schr\"odinger equation with a source term (dipole operator acting on the ground-state wave function) using the hyperspherical harmonics (HH) method~\cite{Grigorenko2009} (solid curve). The rising edge of the peak at $E^*\sim3$~MeV is reproduced correctly, while for higher energies a large discrepancy is visible, despite the fact that the measured spectrum is still dominated by the 2$n$ decay. Finally, a five-body ($^4$He+4$n$) cluster orbital shell model (COSM) was employed using the complex-scaling method to describe many-body unbound states of $^8$He~\cite{Myo2022}. Different excitation modes were found to coexist in the dipole strength distribution (dotted-dashed curve) - below 10~MeV the strength comes from single-particle excitation, $^7$He+$n$, leading to sequential decay into $^6$He+2$n$. At around 13~MeV, collective excitations of four neutrons with strong configuration mixing are enhanced, indicating a possible soft dipole mode. 

\begin{figure}[htbp!]
	\includegraphics[width=\linewidth]{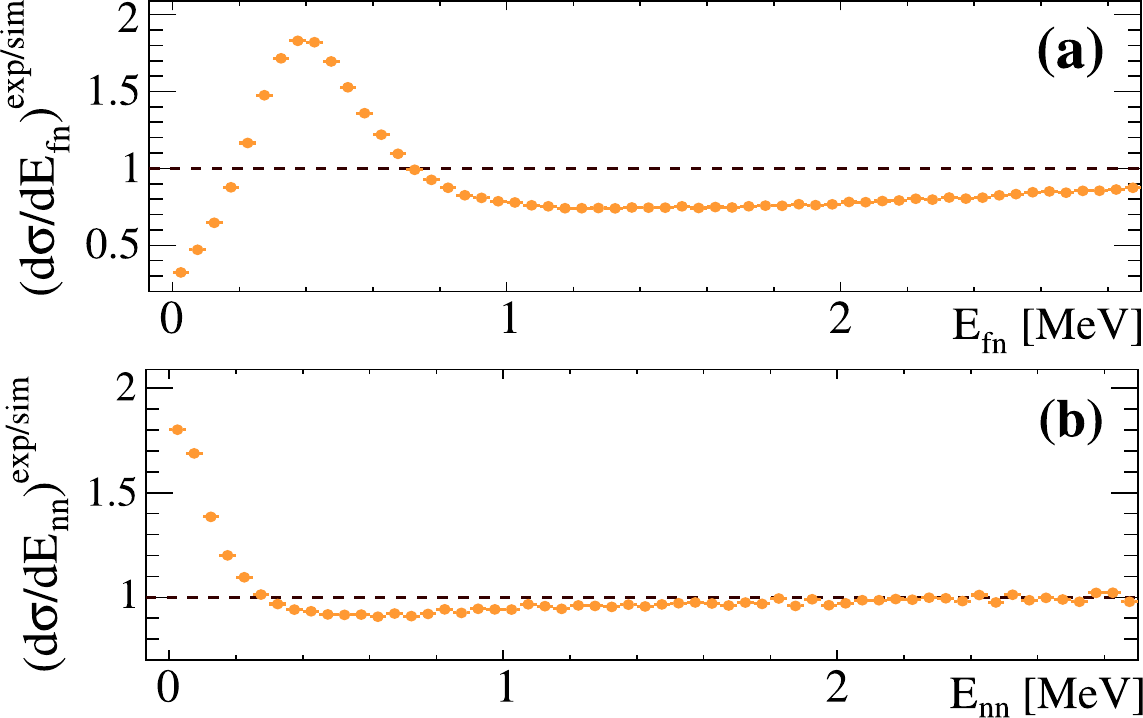}
	\caption{ (a) Fragment-neutron ($fn$) two-body relative energy ($E_{fn}$) for  $^8$He$\rightarrow$$^6$He+2$n$ breakup channel. Shown is the measured spectrum divided by a simulated one assuming phase-space decay. (b) Same for the neutron-neutron sub-system.   } 
\label{fig:corr}
\end{figure}

Correlations in the final state can be investigated by considering the two-body sub-systems. For the $^6$He+2$n$ decay channel this corresponds to the neutron-neutron ($nn$) and fragment-neutron ($fn$) and their corresponding two-body relative energies $E_{nn}$ and $E_{fn}$. For the latter, two possible combinations are taken into account with either of the two neutrons. The two-body correlations are revealed by dividing the measured spectra by simulated ones. Simulations of the three-body ($^6$He+2$n$) system were performed with the differential cross section spectrum as an input, and modeling the breakup after the reaction using a phase-space decay. The measured and simulated data are normalized to each other, such that their division shows deviations of the experimental data from a pure phase-space decay. 

Figure~\ref{fig:corr} (a) shows the $E_{fn}$ relative-energy spectrum, with a clear enhancement at the ground-state energy of the $^7$He resonance. This corresponds to the sequential decay of the three-body system through the resonant state, thus deviating from a direct phase-space decay. In the $nn$ sub-system (panel (b)), low relative energies are enhanced as a result of the $nn$ interaction, reflecting the di-neutron virtual state.  

\begin{figure}[htbp!]
	\includegraphics[width=\linewidth]{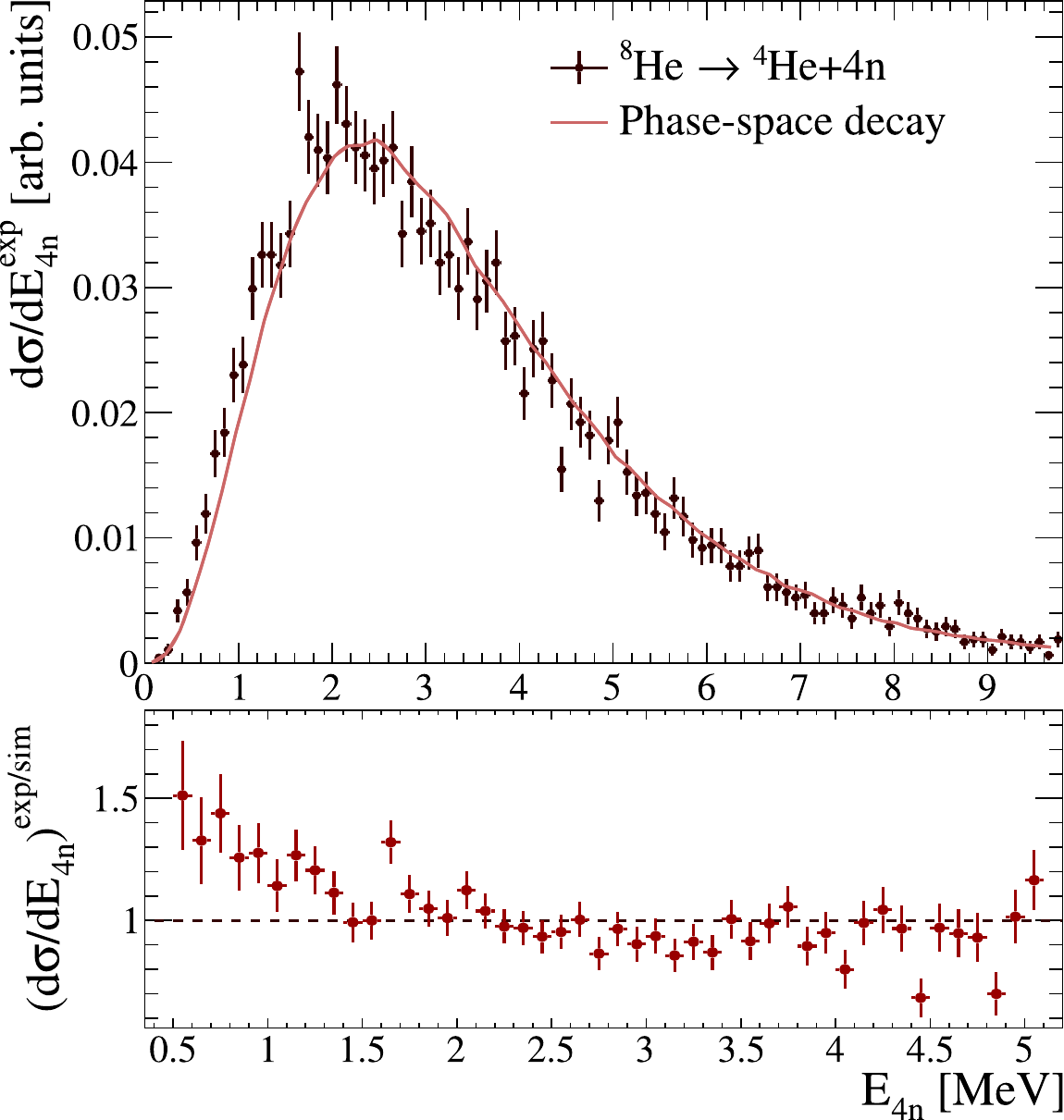}
	\caption{ Top: 4$n$ relative-energy spectrum ($E_{4n}$) for $^8$He$\rightarrow$$^4$He+4$n$ breakup channel. The red curve represents a simulated distribution assuming 5-body phase-space decay. Bottom: measured $E_{4n}$ distribution divided by the simulated one. } 
\label{fig:E4n}
\end{figure}

For the $^4$He+4$n$ decay channel, we study the 4$n$ sub-system to get insight on correlations of the four neutrons. This is of particular interest, following the recent observation of a near-threshold ($\sim$2~MeV) structure in the 4$n$ system via an $\alpha$-knockout reaction from $^8$He~\cite{Duer2022}. Figure~\ref{fig:E4n} (top panel) shows the 4$n$ relative-energy spectrum ($E_{4n}$) in comparison with a simulated phase-space decay, where the bottom panel presents the ratio between them. No indication for a 4$n$ final-state interaction is seen. This is most likely related to the fact that the 4$n$ decay sets in only at rather high excitation energy, as discussed above, preventing a pronounced 4$n$ low-energy correlation. The phase-space decay seems to overall describe the data rather well, while deviations at low relative energy may hint of $nn$ correlations.

{\it Summary and conclusion}.--- In this Letter, we present the dipole response of the most neutron-rich bound nucleus $^8$He up to an excitation energy of 15~MeV as extracted from a measurement of its Coulomb excitation cross section at an energy of $\sim$185~MeV/nucleon. The excitation spectrum has been reconstructed from an exclusive measurement of the residues in the final state using the invariant-mass method, including for the first time besides the $2n$ channel the 5-body decay into $^4$He$+ 4n$. The dipole strength distribution shows a peak at around 3~MeV followed by a broad continuum up to 15~MeV. The dominant decay channel is thereby the $^6$He$+2n$ channel surprisingly even for excitation energies well above the $4n$ threshold, pointing to a dominant di-neutron correlation in the dipole continuum structure. This is supported by the observation of the 2-body correlations in the decay, where strong correlations at the $^7$He resonance energy and the $nn$ virtual state at low energy are observed. In contrast, no strong correlation among the 4 neutrons in the $4n$ channel has been observed, likely related to the fact that the $4n$ channel covers relatively large excitation energies.  

The experimental dipole strength distribution has been compared to an ab initio LIT-CC calculation and to a phenomenological three-body model. The ab initio theory shows good agreement with experiment for excitation energies above 5~MeV, but fails to reproduce the soft-dipole peak at 3~MeV, while the phenomenological three-body model reproduces well the low-energy peak, but misses the higher-lying dipole strength. For the former, a possible reason may lie in missing correlations, while one does not expect the phenomenological cluster model to be valid in the region where the $^4$He core is excited, which is well explained by the ab initio theory. More theoretical efforts need to be devoted in the future to fully describe  the dipole strength distribution of extremely neutron-rich drip-line nuclei.

This work was supported by the Deutsche Forschungsgemeinschaft (DFG, German Research Foundation) through Project-ID 279384907 - SFB 1245 and  Project ID 39083149 PRISMA${}^+$ EXC 2118/1, the GSI-TU Darmstadt cooperation agreement, by the German Federal Ministry of Education and Research - BMBF project number 05P15RDFN1. This work was also supported by the U.S. Department of Energy, Office of Science, Office of Nuclear Physics, under the FRIB Theory Alliance award DE-SC0013617; by the U.S. Department of Energy, Office of Science, Office of Advanced Scientific Computing Research and Office of Nuclear Physics, Scientific Discovery through Advanced Computing (SciDAC) program (SciDAC-5 NUCLEI). C.A. Bertulani acknowledges support from the U.S. Department of Energy, Office of Science, award DE-SC0026074. N.L.A, F.D, J. G., F. M. M., M.P. and N. A. O. acknowledge partial support from the Franco-Japanese LIA-International Associated Laboratory for Nuclear Structure Problems as well as the French ANR-14-CE33-0022-02 EXPAND. S.CP. acknowledges support by  UK STFC under contract numbers ST/P003885/1 and ST/L005727/1 and the University of York Pump Priming Fund. This work was partially supported by the Institute for Basic Science (IBS-R031-D1) in Korea. P. L. acknowledges support from the National Natural Science Foundation of China (Grant No. 12405147). T. N. and Y. K. acknowledge support from JSPS KAKENHI Grants No. JP24H00006 and No. JP18H05404. F.B. and S.B. thank G.~Hagen for providing access to the nuclear coupled-cluster code developed at ORNL (NUCCOR). 

This manuscript has been authored in part by UT-Battelle, LLC, under contract DE-AC05-00OR22725 with the US Department of Energy (DOE). The US government retains and the publisher, by accepting the article for publication, acknowledges that the US government retains a nonexclusive, paid-up, irrevocable, worldwide license to publish or reproduce the published form of this manuscript, or allow others to do so, for US government purposes. DOE will provide public access to these results of federally sponsored research in accordance with the DOE Public Access Plan (http://energy.gov/downloads/doe-public-access-plan).

\bibliographystyle{unsrt}
\bibliography{bib}

\end{document}